\documentclass{ws-procs9x6}

\def\bsax{{\it Beppo}SAX}
\def\sw{{\it Swift}}
\def\ama{$E_{\rm p}-E_{\rm iso}$}
\def\ghi{$E_{\rm p}-E_{\gamma}$}

\def\nfn{$\nu F_{\nu}$}

\def\bb{Black Body}

\begin{document}

\title{GRB~060218 and the outliers with respect to the
\ama correlation}

\author{G. Ghirlanda \& G. Ghisellini}

\address{Osservatorio Astronomico di Brera\\
Merate, I-23807, via E. Bianchi 46\\
E-mail: giancarlo.ghirlanda@brera.inaf.it}

\begin{abstract}
  GRB~031203 and GRB~980425 are the two outliers with respect to the \ama\ 
  correlation of long GRBs. Recently \sw\ discovered a nearby extremely long
  GRB~060218 associated with a SN event. The spectral properties of this
  bursts are striking: on the one hand its broad band SED presents both
  thermal and non--thermal components which can be interpreted as due to the
  emission from the hot cocoon surrounding the GRB jet and as standard
  synchrotron self absorbed emission in the GRB prompt phase, respectively; on
  the other hand it is its long duration and its hard--to--soft spectral
  evolution which make this underluminous burst consistent with the \ama\ 
  correlation of long GRBs. By comparing the available spectral informations
  on the two major outliers we suggests that they might be twins of 060218
  and, therefore, only apparent outliers with respect to the 
  \ama\ correlation. This interpretation also suggests that it is of primary
  importance the study the broad band spectra of GRBs in order to monitor
  their spectral evolution throughout their complete duration.
\end{abstract}


\bodymatter

\section{The peak spectral energy -- isotropic energy in GRBs}

\begin{center}
\begin{figure}
\psfig{file=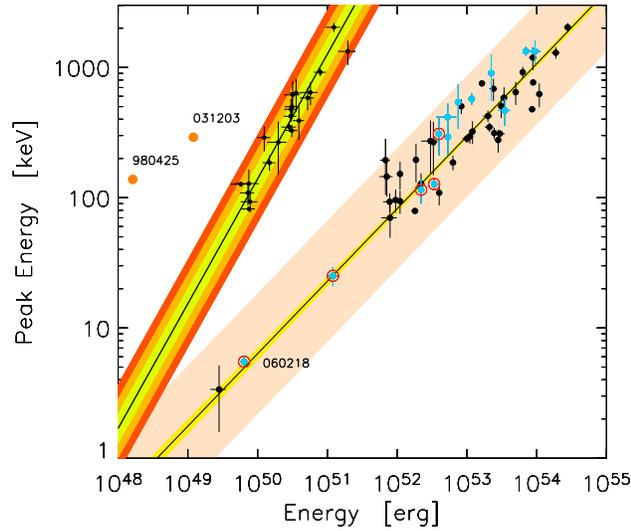,width=3.5in}
\caption{Correlation between the \nfn\ peak spectral energy and the
   isotropic energy (on the left side of the plot) defined with 49
   GRBs (updated to 15 Sept. 2006).The blue points represent the 15
   GRBs added since 2005 (i.e. in the \sw\ ``era'') and the 5 events
   whose peak energy was measured by \sw\ are shown with red--circled
   blue points. The {\it outliers} (GRB~980425 and GRB~031203) are
   shown. On the left side of the plot it is shown the \ghi\
   correlation.}
\label{aba:fig1}
\end{figure}
\end{center}

Long--duration Gamma Ray Bursts (GRBs) present a correlation between
the peak energy of their \nfn spectra ($E_{\rm peak}$) and their
isotropic equivalent energy ($E_{\rm iso}$) emitted during the prompt
phase \cite{ama}. This correlation (presented in Fig.\ref{aba:fig1})
has been updated since its discovery \cite{ama} by adding more than 38
GRBs with measured redshifts and well constrained spectral properties
\cite{ghi,lam,ama1}. For a subsample of these events it was also
possible to estimate their jet opening angles by measuring the jet
break time of their (optical) light curves. The correction of the
isotropic energy for the collimation factor led to the discovery of a
very tight (i.e. the \ghi\ , \cite{ghi,nav}) correlation which has
been used to make GRBs standard candles (\cite{ghi1,fir}). However,
since the discovery of these correlations, GRB~980425 and GRB~031203
resulted inconsistent with them.
GRB~980425 and 031203 are associated with a nearby SN event (at
$z=0.0885$ and $z=0.106$, respectively). However, there are at least
three events which obey the \ama\ correlation and are associated with
a SN event (030329, 021211 and 060218). Among these the most recently
discovered (060218, Campana et al. 2006) could guide us towards the
understanding of the nature of the two outliers. 

\begin{center}
\begin{figure}
\psfig{file=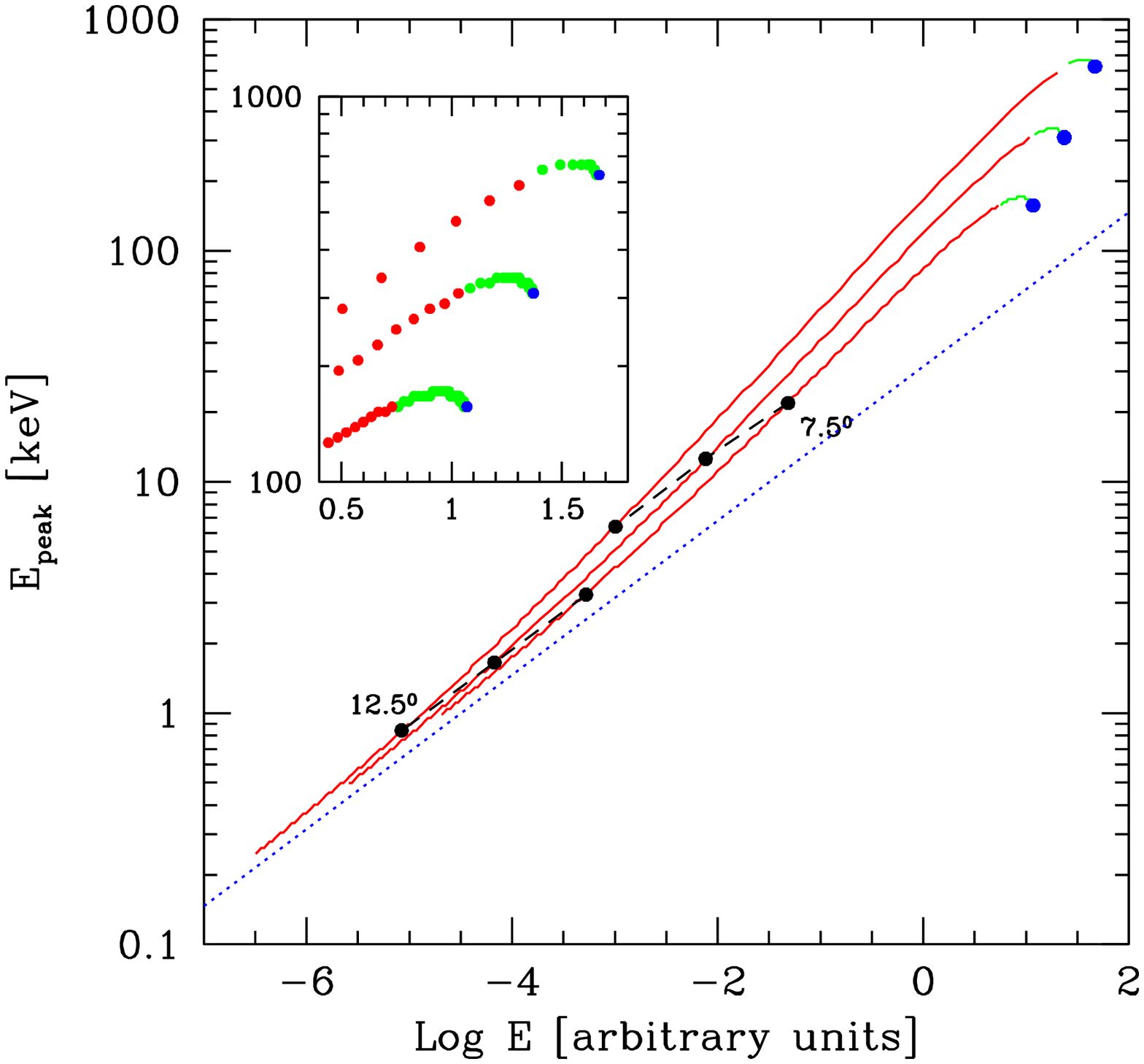,width=2.2in}
\psfig{file=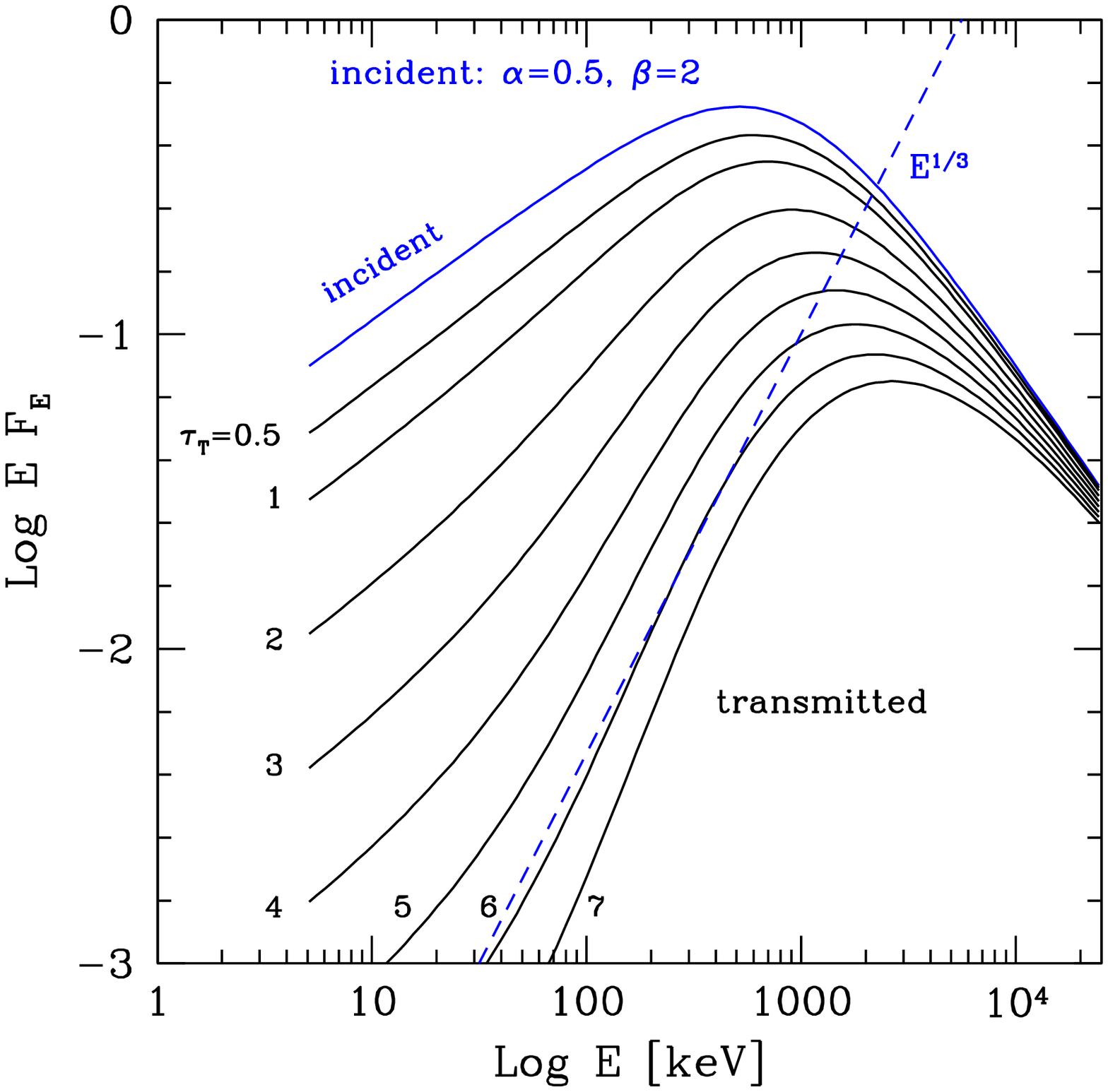,width=2.2in}
\caption{{\bf Left:} The peak of the observed spectrum $E_{\rm peak}$
as a function of the time integrated flux $E$.  Both depend on the
viewing angle $\theta_{\rm v}$.  In the insert we show a zoom for
small viewing angles, within (green dots) and outside $\theta_{\rm
j}$.  We assumed $\theta_{\rm j}=5^\circ$, $0<\theta_{\rm
v}<20^\circ$, $\alpha_1=0.5$, $\alpha_2=2$.  The dotted line, shown
for comparison, has a $1/3$ slope.  The three lines have $\Gamma=50$,
100 and 200, and $E^\prime_{\rm peak}=1.25$ keV.  Black points,
connected with dashed lines, correspond to the same viewing angle
$\theta_{\rm v}=(7.5,12.5)$ for the three different choices of
$\Gamma$. {\bf Right:} Transmitted spectrum for different values of
the Thomson optical depth $\tau_{\rm T}$, as labeled.  The incident
spectrum has $E_{\rm peak}=511$ keV, $\alpha=0.5$ and $\beta=2$.}
\label{aba:fig2}
\end{figure}
\end{center}


It has been proposed \cite{ram,eic} that the two outliers could be
normal GRBs observed off axis (with a typical viewing angle twice
their jet opening angle). In this scenario we can reconstruct the true
energetic and peak energy of these two events if they were observed on
axis \cite{ghis} by correcting for the de--beaming effect
(Fig.\ref{aba:fig2} left panel). It turns out that the two outliers
should be the most luminous events in the population of bursts though
being the closest (980425 is the record--holder) GRBs ever detected.

An alternative possibility is that these two bursts appear
underluminous because their radiation is highly absorbed by material
located in their vicinity (as proposed by \cite{bra,bar}). The
spectrum produced by the central source is modified by the scattering
screen (Fig.\ref{aba:fig2} right panel): for increasing optical depths
the transmitted spectrum has a harder low energy component and a
harder peak energy (with respect to the incident spectrum) due to the
energy dependent Klein--Nishina absorption. In this scenario the two
outliers would require a scattering material of $\tau$ between 6 and 8
to become consistent with the \ama correlation.

\section{GRB~060218: a long burst with a peculiar SED}

\begin{center}
\begin{figure}
\psfig{file=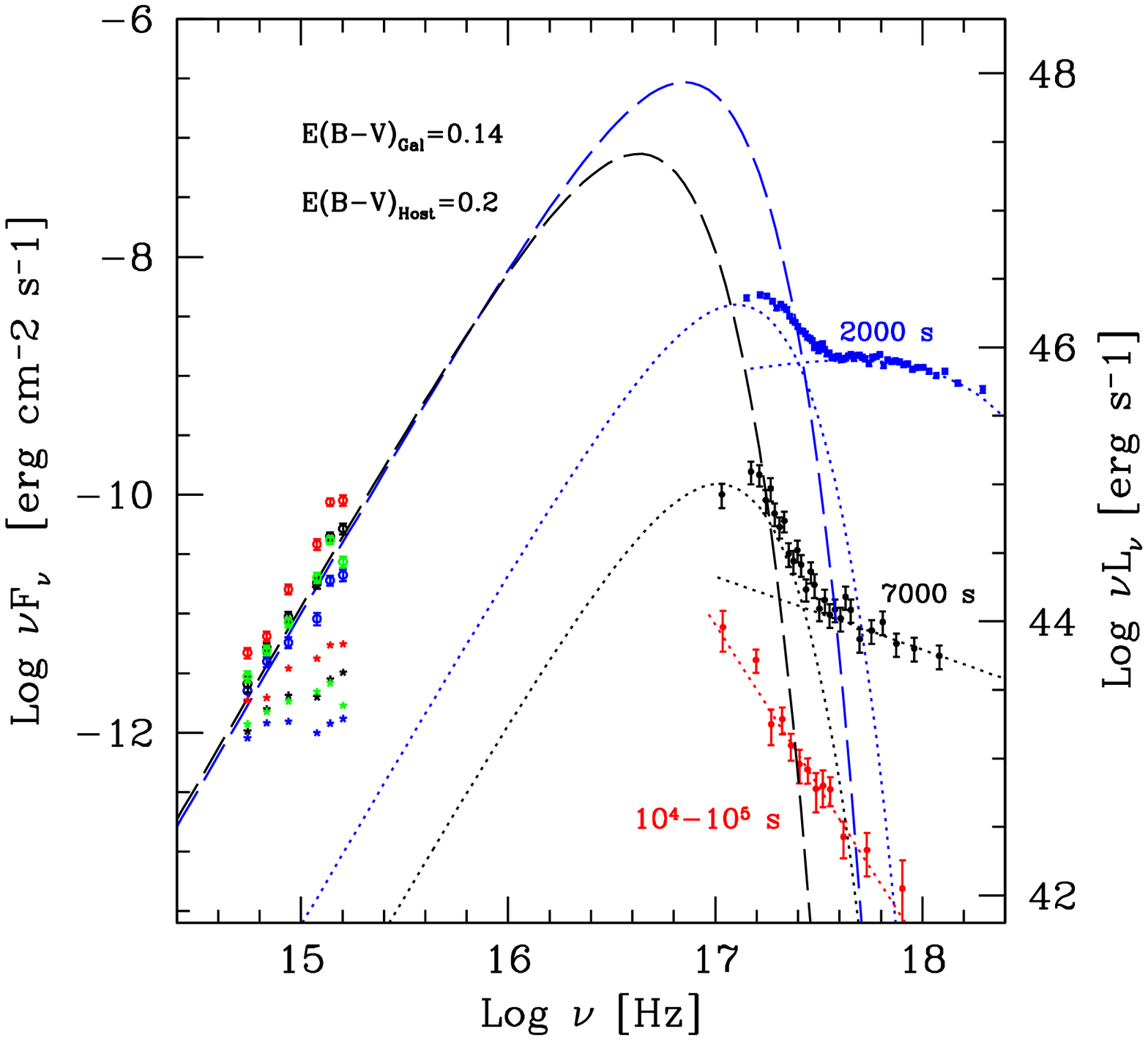,width=2.2in}
\psfig{file=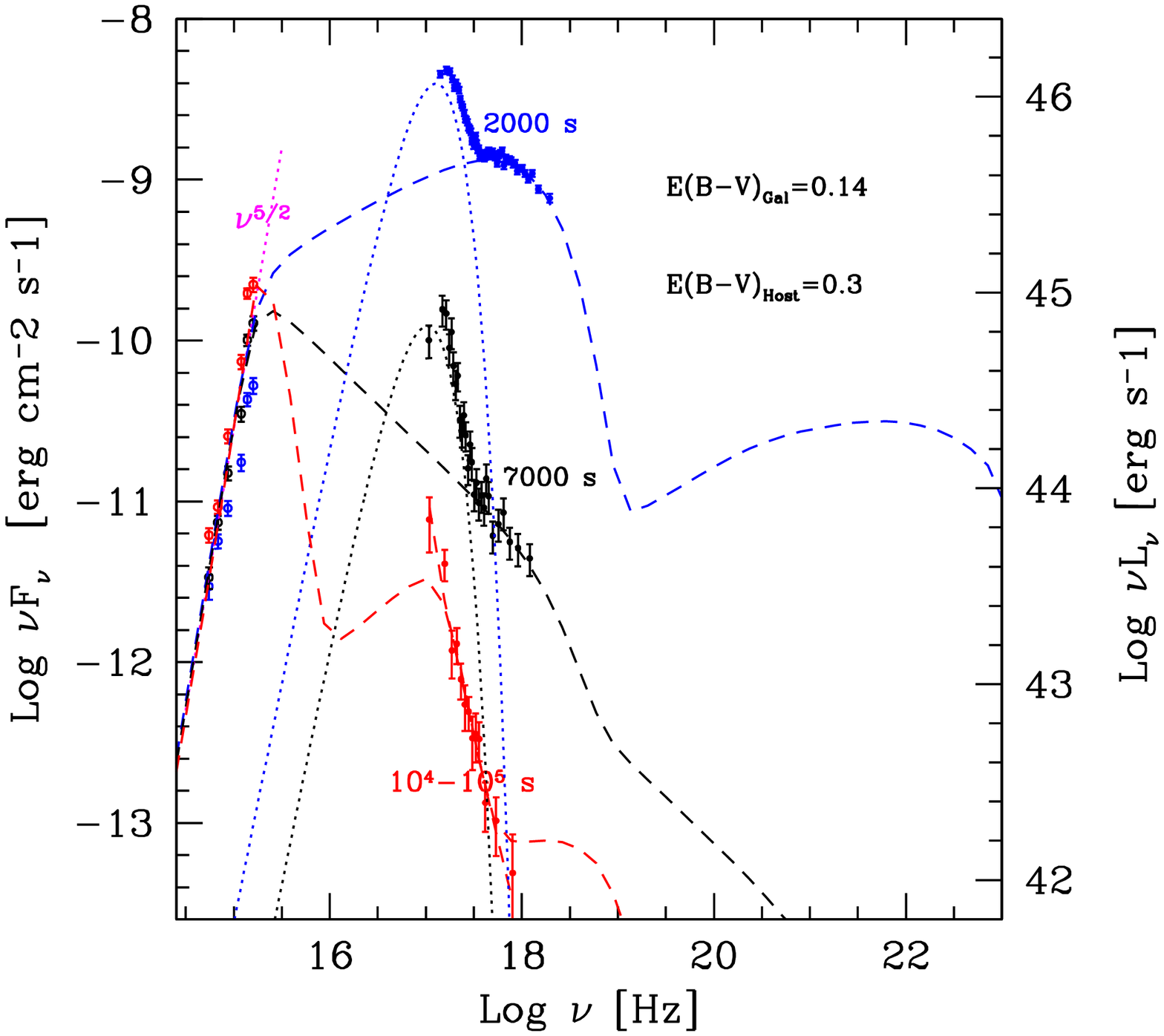,width=2.2in}
\caption{{\bf Left:} The SED of GRB 060218 at different times.  Blue:
2000 s (integrated for $\sim 400$ s for the X--ray); black: 7000 s
(integrated for $\sim$ 2500 s); red: 40,000 s; green: $1.2\times 10^5$
s (only UVOT data are shown).  The opt--UV data are taken from C06
while the X--ray data have been re-analysed by us.  The optical--UV
data lie above the blackbody found by fitting the X--ray data (dotted
lines).  Instead, the opt--UV data seem to identify another \bb
component (long-dashed lines) which is inconsistent with the X--ray
data at the same epochs.  Small crosses without error bars are UVOT
data not de--absorbed.  De--absorbed data [with a galactic
$E(B-V)=0.14$ plus a host $E(B-V)=0.2$] are shown with error
bars. {\bf Right:} The SED of GRB 060218 at different times, as in
Fig. 1, but with the optical UV points de--reddened with $E(B-V)_{\rm
host}=0.3$ instead of 0.2 This produces an opt--UV spectrum $\propto
\nu^{5/2}$.  We also show the SSC model, discussed in the text, for
the 3 SEDs for which we have simultaneous UVOT, XRT data (i.e. at
2000, 7000 and $\sim 10^4$--$10^5$ seconds after trigger). }
\label{aba:fig3}
\end{figure}
\end{center}

GRB~060218 ($z=0.033$, \cite{mir}), associated to SN 2006aj \cite{maz}
is a long duration event ($>$3000 s) detected by \sw\ BAT and followed
with a few hundred seconds delay by the XRT and UVOT telescopes
on--board \sw\ \cite{cam}. The broad band Optical to X--ray
SED of GRB 060218 (Fig.\ref{aba:fig3} left panel) presents some
interesting features: (i) a (steady) thermal component in the X--ray
with typical temperature of $\sim$0.2 keV and a total energy of
$\sim10^{49}$ erg; (ii) a non--thermal X--ray component softening with
time and (iii) a (steady) opt-UV spectrum which is well described by
the Rayleigh--Jeans tail of a black body.

The presence of a \bb\ component has been interpreted \cite{cam} as
the SN shock breakout (SNSB) emission, which has never been observed
before.

As shown in Fig.\ref{aba:fig2} (left panel), the opt--UV spectrum
lies above the extrapolation of the X--ray Black Body. Instead, a single \bb\ 
(whose Rayleigh--Jeans tail matches the opt-UV data) is inconsistent with the
X--ray spectrum. Moreover, the latter possibility requires that the \bb\ 
luminosity is $10^{48}$ erg/s. Considering the exceptional duration of this
burst (i.e. $>10^{3}$ s) this would imply that, if this is the energy produced
by the subrelativistic SN shock breakout, it would exceed the total kinetic
energy of the SN (i.e. $\sim10^{51}$ erg) estimated from the late time optical
spectroscopy \cite{maz}.

On the other hand it might still be possible that either the X--ray or
the opt-UV \bb\ component are the SNSB. Nonetheless, in the first case
the velocity of the emitting material ($v=(L_{BB}/4\pi t^2 \sigma_{r}
T_{BB}^{4})^{1/2}$) is very low ($\sim$3000 km/s) compared to the
velocity derived from the optical spectroscopy ($\sim$20.000 km/s -
\cite{pia}). In the second case, instead, the \bb\ temperature
should not be much above the UV frequency (to limit the total
energetic) but the velocity derived (if the SN exploded simultaneously
to the GRB) is larger than $c$. Moreover, detailed numerical modeling
of the SNSB (Li 2006) predicts lower luminosity and duration and
larger temperatures for the X--ray emission of the SNSB than what
observed.

We have instead proposed \cite{ghis1} that the opt--UV spectrum and
the non--thermal contemporaneous X--ray emission of GRB 060218 can be
well fitted with a synchrotron--self--absorbed model
(Fig.\ref{aba:fig3} right panel): in this case the self--absorption
frequency falls just above the opt-UV band. The X--ray \bb\ component,
instead, is the thermal emission from the hot cocoon surrounding the
jet (e.g. \cite{ram1,fan}).

\begin{center}
\begin{figure}
\psfig{file=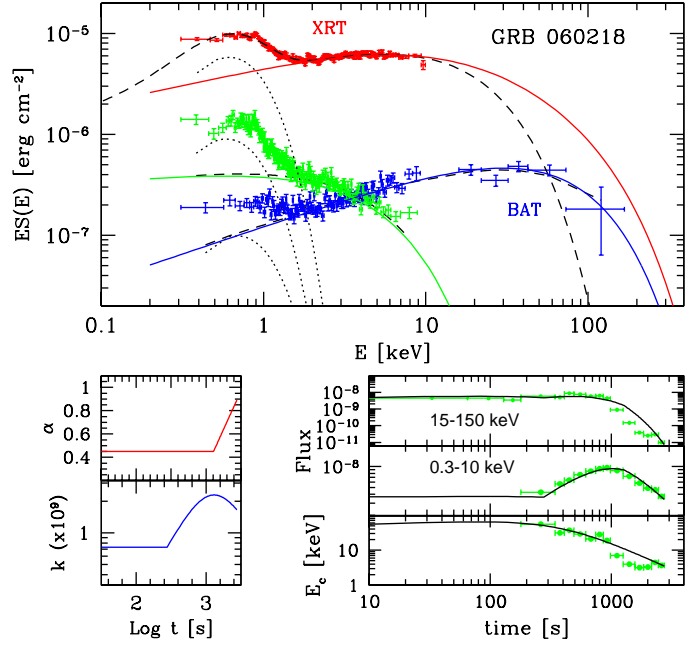,width=4.0in}
\caption{Top panel: Spectra of GRB 060218 for different time--bins: i)
entire duration (top); ii) [159--309 s] (rising spectrum with also BAT
data) iii) [2456--2748 s] (soft spectrum).  We plot $ES(E)$ vs $E$,
$S(E)$ being the fluence.  Dotted lines indicate the blackbody
component, not considered for the spectral evolution, and long--dashed
lines represents the best fit obtained from the analysis of the data.
Continuous lines show the results of our proposed modeling.  Left
bottom panel: assumed behaviour of the normalisation $K$ and energy
spectral index $\alpha$.  Right bottom panel: light curves in the BAT
(15--150 keV) and XRT (0.3--10 keV) range, and evolution of $E_{\rm
c}$.  The flux in the 0.3--10 keV is the (de--absorbed) flux of the
cut--off power law component only: we have subtracted the blackbody
component from the total flux.  Continuous lines are the results of
our modelling.}
\label{aba:fig4}
\end{figure}
\end{center}

\subsection{The spectral evolution of GRB 060218}

The most striking spectral property of GRB 060218 is that its spectrum
evolves from the hard BAT energy band to the soft XRT band. For this
reason the time--integrated spectrum of this burst has a peak energy
in the soft X--ray band at 5 keV. Considering its relatively low
luminosity (i.e. $\sim 10^{49}$ erg - similar to that of the two
outliers), its low $E_{\rm peak}$ is what makes it consistent with the
\ama\ correlation. In Fig.\ref{aba:fig4} we show two spectra
(corresponding to the initial and the final emission of the burst) and
its time integrated spectrum. The other panels show the fit with a
model that reproduces the spectral evolution and the light curve in
the 0.2-10 keV and 15-150 keV band and the time evolution of the
$E_{\rm peak}$.

In particular the presence of nearly simultaneous observations of the
burst prompt emission by the XRT instrument on--board \sw\ was the key
to classify this burst as being consistent with the \ama\ correlation:
in fact, if only BAT measured its spectrum, we would have classified
this event as the third outlier with respect to the \ama\ correlation,
with $E_{\rm peak}\sim 100$ keV and $E_{\rm iso}\sim 7\times 10^{48}$
erg.

\begin{center}
\begin{figure}
\psfig{file=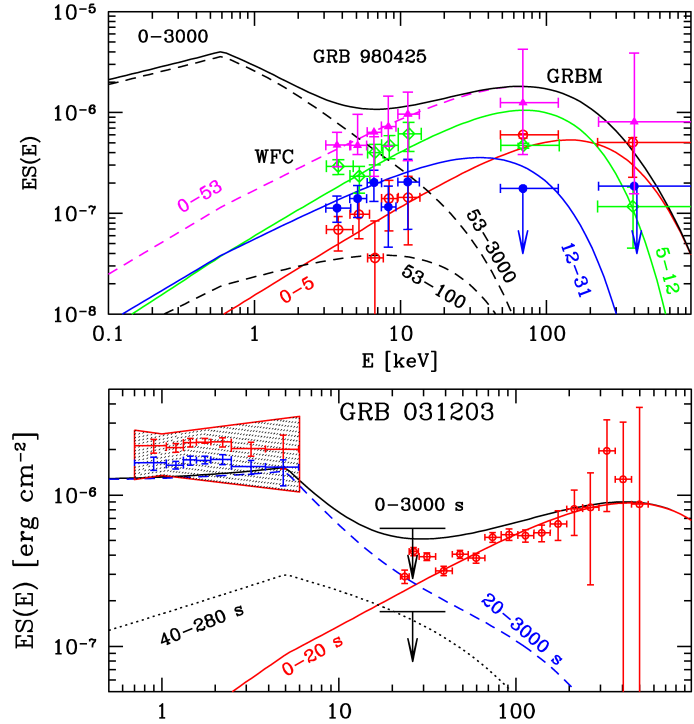,width=4.0in}
\caption{{\it Top panel}: spectral evolution of GRB~980425. The
   data are from \bsax (WFC: 2--28 keV and GRBM: 40--700 keV adapted
   from Frontera et al. 2000). The model fits (lines) are obtained
   with the same model used for GRB~060218 by simultaneously fitting
   the light curves and the available spectra of GRB~980425. {\it
   Bottom panel}: spectral evolution of GRB~031203. In this case the
   late time spectrum should produce a considerable flux in the X--ray
   band to be consistent with the observed evolution of its dust
   scattering halo \cite{tie}.}
\label{aba:fig5}
\end{figure}
\end{center}

\section{GRB031203 \& GRB980425 become mainstream}

We have verified if GRB 031203 and GRB980425 have a spectral evolution
consistent with that of 060218 (i.e. hard to soft). If this is the
case it is possible that their soft late--time emission went
undetected in the soft X--ray instruments on board Integral and
BeppoSAX, which detected these two events. Interestingly, GRB 031203
(Fig.\ref{aba:fig5}) produced a spectacular dust scattering halo
(observed with XMM--Newton) which evolved in time. The spectral flux
responsible of the halo should have had a fluence similar to that of
the prompt emission detected by Integral. This has two effects: on the
one side the total energy is larger than that measured from the
Integral spectrum alone while, on the other side, the peak energy of
the time integrated spectrum is in the X--ray band. This two effects
combined make GRB 031203 consistent with the \ama correlation.

In the case of GRB 980425 our model predicts a considerable long
duration of the burst with a spectrum peaking in the soft X--ray band
at late times. Unfortunately there are no data confirming this
possibility (as opposed to the case of 031203) and we are therefore
forced to assume that this burst lasted more than what the WFC
on--board BeppoSAX could monitor.

\section*{Acknowledgments}
We are grateful to F. Tavecchio, C. Firmani, Z. Bosnjak for fruitful
collaborations.

\end{document}